\documentclass{egpubl}
\usepackage{eurovis2025}

\ConferencePaper        
\SpecialIssuePaper         

\CGFccby

\usepackage[T1]{fontenc}
\usepackage{dfadobe}

\usepackage{cite}  
\BibtexOrBiblatex
\electronicVersion
\PrintedOrElectronic
\ifpdf \usepackage[pdftex]{graphicx} \pdfcompresslevel=9
\else \usepackage[dvips]{graphicx} \fi

\usepackage{egweblnk}

\usepackage{xcolor}
\usepackage{tcolorbox}
\usepackage{fontawesome5}

\newcommand{\toolname}{\textsc{DashGuide}}

\newcommand{\qte}[2]{\textit{``#1''}~(#2)}

\newcommand{\cre}[1]{\textcolor{black}{#1}}
\newcommand{\cree}[1]{\textcolor{black}{#1}}

\title[DashGuide]%
      {DashGuide: Authoring Interactive Dashboard Tours for Guiding Dashboard Users}

\author[N. Hoque \& N. Sultanum]
{\parbox{\textwidth}{\centering N. Hoque$^{1}$\orcid{0000-0003-0878-501X}
        and N. Sultanum$^{2}$\orcid{0000-0001-8608-1427} 
        }
        \\
{\parbox{\textwidth}{\centering $^1$University of Iowa, Department of Computer Science, USA\\
         $^2$Tableau Research, USA
       }
}
}

\begin{document}

\teaser{
\centering
  \includegraphics[alt={A teaser image with five subfigures. The first figure shows a full dashboard with three hand cursors positioned in three corners of the dashboard. Each of the other four subfigures shows a partial view of the dashboard with some text placed on top of it.},
  width=0.95\linewidth]{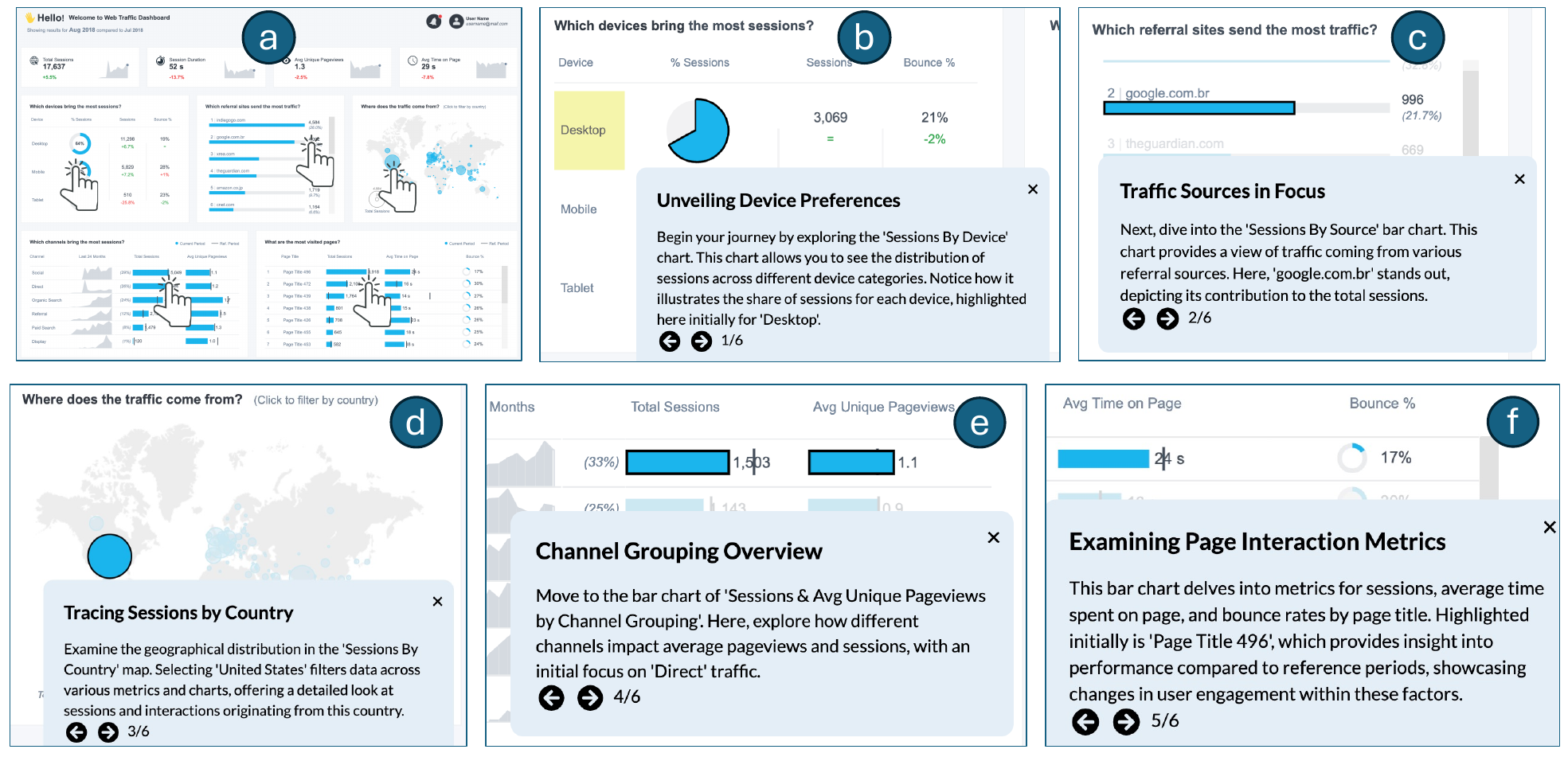}
  \caption{\textbf{Overview of \toolname{}.} a) The dashboard in this example visualizes different metrics from web traffic data (Sourced from Tableau Public \href{https://public.tableau.com/app/profile/pradeepkumar.g/viz/WebTrafficDashboardRedesign/Cockpit}{[link]}). To create a guided tour for this dashboard, the dashboard author first selects ``Dashboard Semantics and Encoding'' as the communication goal  for the tour (not shown) and then performs several consecutive click selections on the dashboard (shown with the hand cursors). \toolname{} automatically generates content for the tour by prompting a large language model (LLM), which uses the communication goal and interactions provided by the author. The tour is then delivered as a step-by-step tour (b-f) where each step plays back the original interaction while placing the title and description for the step as an overlay near the location of interaction. The author can refine the content manually, provide further instructions to the LLM, or add or remove steps from the tour. }
\label{fig:teaser}
}

\maketitle
\begin{abstract}
  \cre{Dashboard guidance helps dashboard users better navigate interactive features, understand the underlying data, and assess insights they can potentially extract from dashboards. However, authoring dashboard guidance  is a time consuming task, and embedding guidance into dashboards for effective delivery is difficult to realize. In this work, 
  we contribute} \toolname{}, a framework and system to support the creation of interactive dashboard \cre{guidance} with minimal authoring input. Given a dashboard and a communication goal, \toolname{} captures a sequence of author-performed interactions to generate guidance materials delivered as playable step-by-step overlays, a.k.a., dashboard tours. Authors can further edit and refine individual \cre{tour} steps while receiving generative assistance.
  \cre{We also contribute findings from a formative assessment with 9 dashboard creators, which helped inform the design of \toolname{}; and findings from an evaluation of \toolname{} with 12 dashboard creators, \cree{suggesting it provides} an improved authoring experience that balances efficiency, }expressiveness, and creative freedom.


\end{abstract}

\section{Introduction}

Visualization dashboards were conceived as tools for independent, self-service data exploration~\cite{DBLP:journals/cga/ToryBFC23}. In practice, however, dashboard users often require \cre{assistance in understanding components and functionality of a dashboard, from operating data filters to understanding the underlying data, interpreting charts, and navigating the dashboard to answer specific analytical questions~\cite{DBLP:journals/tvcg/CenedaGMMSST17, DBLP:journals/cgf/CenedaGM19, DBLP:conf/chi/KwonL16}.} To this end, dashboard authors routinely provide \textit{guidance} content to help \cre{these users,} particularly those new to a dashboard~\cite{DBLP:journals/cgf/DhanoaWHSGS22, tvcg/choe, DBLP:journals/vi/StoiberCWSGSMA22}.



\cre{Common forms of dashboard guidance 
include embedded text instructions and standard UI elements such as pop-ups and tooltips~\cite{DBLP:journals/tvcg/CenedaGMMSST17, DBLP:journals/vi/StoiberCWSGSMA22, DBLP:journals/tvcg/SultanumText2024}.} While such guidance is relatively easy to author and often supported by authoring tools like Tableau, they tend to lack user engagement and may be skipped, missed or ignored by users~\cite{DBLP:journals/tvcg/SultanumText2024, DBLP:conf/vinci/StoiberWGSSA21}. More recently, researchers have proposed interactive \textit{dashboard tours} to improve user engagement and show diverse analytical capabilities of a dashboard~\cite{dhanoad, DBLP:conf/chi/NetworkNarratives}. These tours \cre{show} usage scenarios, interactions, and insights in a step-by-step format, with each step incorporating a combination of content elements such as text descriptions and data highlights. Although recent \cre{work provides the tooling to instrument such tours~\cite{dhanoad}, authors still need to manually draft the content itself (e.g., text descriptions and images), which demands significant time and effort and offers limited scalability to supporting multiple tours with diverse usage scenarios.}

\cre{In view of these challenges, our work contributes: (a) a \textit{framework and system, }\toolname{}, to support dashboard creators in authoring dashboard tours and content more easily; (b) \textit{findings from a formative assessment} that guided design and development of \toolname{}; and (c) \textit{findings from a qualitative user evaluation of \toolname{}} that underscore the value of our approach.}

\cre{Our user-centered design journey began with formative interviews with 9 dashboard practitioners. Interview findings point to time and effort being the biggest obstacles to crafting engaging and comprehensive guidance. While dashboard creators favor dynamic forms of guidance such as dashboard tours, technical limitations result in guidance being delivered in mostly static form (e.g., technical documentation external to the dashboard). Interviews also informed a range of communication goals for dashboard guidance, from demonstrating coordinated views to presenting key insights and explaining chart encodings.}

Based on this feedback, we designed \toolname{}, our solution for \cre{easier authoring of} interactive dashboard tours. 
To expedite the authoring process while accommodating author oversight and communication goals, \toolname{} proposes an \textit{interaction-first} approach. In this approach, a creator can quickly express their intent by choosing a high-level communication goal and then interacting with components of interest in the dashboard. \toolname{} captures the interaction, compiles a sequence of key demonstration segments, and links interacted dashboard components and corresponding interactive actions with the respective segments. A Large Language Model (LLM) is then used to generate relevant content for each key segment, steered by the author's informed communication goal. The dashboard tour is then rendered as a step-by-step guide where each step replays the corresponding interactions for the step, along with an explanatory title and description placed near the interacted components as overlays (\autoref{fig:teaser}). After an initial tour is created, \toolname{} provides several edit and refinement features, including inserting new steps and regenerating content. 





We evaluated \toolname{} in a user study with 12 dashboard creators where we asked participants to author dashboard tours for their own (or publicly available) dashboards. \cree{Qualitative} findings \cree{suggest} that \toolname{} provided \cree{them with meaningful authoring support}, and that the proposed interaction-first framework is a \cree{potentially} promising approach for authoring dashboard tours. Our investigation also revealed open challenges around guidance authoring that future embodiments of \cree{our framework} may help solve. 
\section{Related Work}
Our work builds upon prior works on tutorial systems in HCI, dashboard guidance, and data tours. We discuss these areas below.

\subsection{Guidance and Tutorial Systems in HCI}



Guidance in HCI broadly encompasses problems and aids associated with learning to use software and hardware~\cite{DBLP:conf/chi/GrossmanF10}. The traditional approach to providing guidance is text documentation~\cite{carroll1990nurnberg}. However, documentation can be difficult to follow~\cite{DBLP:journals/cacm/Rettig91}, making them less accessible to users. 
A long line of HCI research has proposed solutions to make guidance more engaging and useful. The general consensus is that users prefer guidance that can show usage scenarios instead of outlining them in a static format~\cite{DBLP:conf/chi/GrossmanF10, DBLP:conf/uist/BergmanCLO05, DBLP:conf/uist/ChiFPE21, DBLP:journals/tog/GrablerALDI09, DBLP:conf/uist/LiGF12}. 
One popular approach is to adopt a step-by-step guided tour to show usage scenarios. Several works have proposed such tours for mobile  applications~\cite{DBLP:conf/uist/0001LCL21, DBLP:conf/uist/ChiFPE21}, desktop and web applications~\cite{DBLP:conf/uist/ChiARDLH12}, and immersive environments (AR/VR)~\cite{DBLP:conf/chi/KumaravelNDH19, DBLP:conf/chi/HuangQWPSCRQ21}.
Several open-sourced libraries (e.g., Intro.js~\cite{introjs}, TourGuide.js~\cite{tourguidejs}, and WebTour.js~\cite{webtourjs}) are also available for creating guided tours. This line of work inspired us to use guided tours that can sequentially show usage scenarios for a dashboard (i.e., dashboard tours).

While guided tours are popular, authoring them can be a challenging task. One prominent approach is to construct the steps for the tour by recording user demonstrations. For example, EverTutor~\cite{DBLP:conf/chi/WangCCHC14} records touch events on a mobile phone to identify the steps for a tour. MixT~\cite{DBLP:conf/uist/ChiARDLH12} features a similar recording process but focuses more on combining images and videos for tutorial steps. Commercial tools such as Scribe~\cite{scribe} also generate steps from user demonstrations. 
This line of work inspired our \textit{interaction-first} approach where an author's demonstration, along with author-provided metadata, is used to determine the steps and narrative of the tour. We then utilize an LLM to generate content for the steps, which the author can further refine. To our knowledge, our work is the first to introduce such an interaction-driven and LLM-based method to author dashboard tours.


\subsection{Characterizing Dashboard Guidance}
\label{sec:dashboard_guidance}



Ceneda et al.~\cite{DBLP:journals/tvcg/CenedaGMMSST17} proposed the most comprehensive model for characterizing dashboard guidance. According to this model, users may need guidance because they either do not know the desired results or the path to reach desired results. The model discusses five possible domains where users may need guidance: data, tasks, VA methods (i.e., charts, widgets, and interactions), users, and infrastructure. 
Finally, the model discussed three levels of guidance: \textit{orienting}, \textit{directing}, and \textit{prescribing}. Orienting involves helping users build and preserve their mental map by exposing patterns, data subsets, insights, and provenance~\cite{DBLP:journals/tvcg/GratzlGLPS14, DBLP:journals/cgf/LexSSPSPG12, DBLP:conf/infovis/KreuselerNS04}. Directing provides recommendations or alternative choices to users. Recommender systems such as VizAssist~\cite{DBLP:journals/vc/BoualiGV16} and Voyager~\cite{DBLP:journals/tvcg/WongsuphasawatM16} fall in this bucket. Finally, Prescribing, the focus of this paper, show usage scenarios and analytic processes of a dashboard through demonstration, often in a step-by-step tour format (i.e., dashboard tour)~\cite{DBLP:journals/cgf/YuLRC10, dhanoad, DBLP:conf/chitaly/StoiberRGMSGGA23}. This type of guidance is often employed to onboard users in a dashboard and reduce learning curves~\cite{DBLP:journals/cgf/DhanoaWHSGS22, dhanoad, DBLP:journals/vi/StoiberMSGPGSA24}.

 \toolname{} falls in the \textit{prescribing} category, as our dashboard tours primarily enable dashboard creators to prescribe usage scenarios to end users.  Our work extends previous works focusing on prescribing in two ways. First, despite previous efforts to characterize prescribing, we still lack an understanding of how dashboard creators author such guidance in practice and what kind of support they require in this task. This work fills that gap through a formative study with dashboard creators. Secondly, this work proposes a system to author dashboard tours, an effective way to prescribe guidance to end users~\cite{DBLP:journals/vi/StoiberMSGPGSA24, dhanoad}. The system includes several components to expedite the authoring process: an interaction-based approach to capture the author's communication goal,  an LLM-supported content generation strategy, and in-situ playback of the dashboard tours as overlays in the dashboard.

 It should be noted that we do not limit the purpose of our dashboard tours to onboarding alone, although prescribing-focused guidance is typically designed for onboarding users~\cite{DBLP:journals/tvcg/CenedaGMMSST17}. This is because we wanted to understand the design space and potential use cases of dashboard tours as guidance beyond onboarding materials. 
 \cree{
 We also note that we have not conducted  studies with end-users in this paper due to our focus on supporting dashboard \textit{authors} prescribe guidance. 
 This is a complex and nuanced task~\cite{DBLP:journals/tvcg/CenedaGMMSST17}, which made us anticipate that we would need extensive engagement with authors (i.e., dashboard creators) to understand their needs holistically and evaluate our solution. Thus, we decided to prioritize engagements with authors while leaving engagements with end-users for future work. }

\subsection{Data Tours, Storytelling, and Guidance}


Data tours \cre{consist of} 
interactive step-by-step data-driven content that walk users through a narrative or story~\cite{DBLP:conf/chi/NetworkNarratives}. From Asimov's ``The Grand Tour''~\cite{asimov}, several use cases of data tours have emerged.  \cre{A} prominent use of data tours is data storytelling~\cite{riche2018data, DBLP:journals/cga/ZhaoE23, DBLP:journals/tvcg/SegelH10}: news outlets such as The New York Times often use step-by-step format for data-driven stories~\cite{nytimes_datatours}, \cre{ plus works such as} VizFlow~\cite{DBLP:conf/chi/SultanumCBL21}, graph comics~\cite{DBLP:conf/chi/BachKHCKR16}, and NetworkNarratives~\cite{DBLP:conf/chi/NetworkNarratives}.

 \cre{Dashboard tours in \toolname{} follow the step-by-step} format of a data tour. We opted to not use the term ``data tour'' as it is widely used in the data storytelling space and may create confusion with our focus on guidance. While not as frequently as for storytelling, prior works have used guided tours as guidance. For example, IBM Cognos Analytics~\cite{ibm_cognos} uses a tour to show different functionalities of the platform. The most closely related work to ours is D-Tour~\cite{dhanoad}. In this work, the authors proposed a system that allows authors to arrange automatically extracted dashboard components into multiple tours with possible shared steps between the tours. While D-Tour offers a flexible \cre{way to} author multiple tours, authors have to manually construct the narrative for the tour and create content for each step. In contrast, \toolname{} expedites authoring by providing content generation assistance that aligns with authors' communication goals.
Our mechanism also simplifies creating the steps and is more suitable for highlighting nuanced insights and scenarios,
as authors can set focus to any data points and components just by interacting with the dashboard. 


Another closely related work is LEVA~\cite{leva}, an LLM-based system that can generate onboarding materials for a dashboard. But contrary to our system, LEVA is fully automated and
dashboard authors have few resources to steer content generation in a way that aligns with their communication goals.
Our system prioritizes \cre{author agency and the content they want to show in guidance by providing an efficient way to capture those goals, while also providing} author-driven tools for content refinement.

\section{Formative Study}
\label{sec:formative}
Our literature review revealed a knowledge gap between our theoretical understanding of prescribing guidance and common practices among dashboard creators in regards to authoring such guidance. To address this gap, we conducted semi-structured interviews with dashboard creators to understand how they currently author guidance focused on prescribing, their challenges, and potential solutions for authoring such guidance.

\subsection{Participants and Methodology}
\label{sec:formative-participants}
We recruited nine participants (P1-P9)  by advertising in a specialized community forum of data analysts and professional dashboard creators. 
Participants were data professionals with occupations \cre{ranging from business intelligence (BI) managers (3/9), developers (5/9), and a data consultant (1/9)\,---\,more details in supplemental materials}. Participants had at least five years of experience producing dashboards, \cre{and were familiar with dashboard guidance authoring in various formats,} such as technical documentation, text overlays, videos, and demos.

We conducted 45-min semi-structured interviews via video-conferencing, divided into three parts. In the first part, we asked participants about their professional background and target users. The second part focused on participant experiences with creating guidance, specially on prescribing guidance to end users. In the third part, participants brainstormed with the research team to outline requirements and design choices for a guidance authoring system. The 1st author conducted the interviews, while the 2nd author took notes and joined discussions. 
We conducted thematic analysis~\cite{braun2006using} of interview transcripts and study notes. Responses were aggregated across participants and recurrent themes were identified under each question. 
We present our findings below, while establishing parallels with other relevant findings in past literature.

\subsection{Findings}
\label{sec:formative-findings}

Participants shared their experiences on creating dashboards for a broad range of audiences and domains. Some supported dashboarding needs of small-to-medium specialized teams in their respective organizations (P3, P7-9); some supported much larger and diverse audiences with wildly varying levels of visual literacy skills (P1, P4, P5); and some worked with a broad range of clients in a consulting capacity (P2, P6). While their experiences are extremely diverse, they shared many of the same challenges and tasks. We compiled their feedback into the following topics.




\subsubsection{Authoring guidance is a nuanced task}
\label{sec:formative-findings-nuanced}
Participants stated that guidance is an integral part of their dashboard authoring practices. The primary reason for providing guidance is to attend to the diverse skills and needs of the target user bases or clients (P1-4, P6, P8-9).  Prior research has also reported the challenge with visual literacy skills as the primary reason for providing guidance~\cite{DBLP:journals/cgf/TanahashiLM16, tvcg/choe, DBLP:journals/tvcg/CenedaCEMTA24}. Participants mentioned that while their dashboards typically have specific analytical purposes, the target user bases may not possess the skills or necessary context to realize those analytical purposes. On that, P2, frames the act of providing guidance as a \textit{translation} exercise: \textit{``We assume people have the same understanding of language or words, which is not true.''} This means that \textit{good} guidance goes beyond a mechanical description of what dashboard components are and what they do, but is rather a nuanced \textit{authoring} exercise: one that requires being attuned to a user's context, picking the right words, and delivering them in an accessible way. This means that, instead of pursuing a fully automated guidance generation solutions, we should instead support authors realize their authoring potential. 




\subsubsection{Interactive and embedded guidance is the gold standard} 
\label{sec:formative-authoring-guidance}
Text annotations (4/9) and on-off overlays (5/9) were some of the most common prescribing guidance formats mentioned by practitioners. These forms of embedded guidance are desirable from a usability and convenience standpoint, as 
\qte{you don't want [users] to leave the dashboard}{P5}. 
A limitation of these formats is that they are almost always static, and therefore, limited in terms of what aspects of the analytical process they can highlight: \qte{Text overlays are my go-to guidance format. I typically have a button in the dashboard that opens an overlay layer with explanatory text on top of the dashboard. However, the overlays typically provide very high-level guidance and I cannot show any analytical process with that}{P2}. Another limitation is that embedded guidance can be very high maintenance, often requiring  third-party software (e.g., Figma) to author and manual labor to keep them updated: \qte{The biggest challenge is updating the overlays. Every time I make any changes to the dashboard, I have to update my Figma design and then transform it to my dashboard}{P6}.
Other forms of guidance include on demo videos (P3-4, P7-8), in-person walkthroughs (P2-3, P7, P9), and phone calls  (P1, P4-5). 

Interestingly, none of the participants spontaneously mentioned dashboard tours to be a common practice for prescribing guidance, despite prior research acknowledging their effectiveness~\cite{dhanoad, DBLP:conf/chi/NetworkNarratives, DBLP:conf/uist/ChiARDLH12}. When asked, all participants stated not knowing how to author tours, but were intrigued by the idea: \textit{``Oh! [tours] would be really cool. I kind of take a similar approach when I create videos, but it takes a lot of time to create just one video. If I can create multiple tours quickly, that would be super helpful'' (P4).} P8 said: \textit{``Tours could be really helpful if they can highlight data points and show interactions. But I don't know how I can do that.''} This confirms that dashboard tours are an appealing prescribing format but remain difficult to realize in practice.



\begin{figure*}
    \centering
    \includegraphics[alt={There is a dashboard with white background on the right. On the left, there is a rectangular block with some text and small icons. A similar rectangular block is in the middle of the image.},
    width=0.85\textwidth]{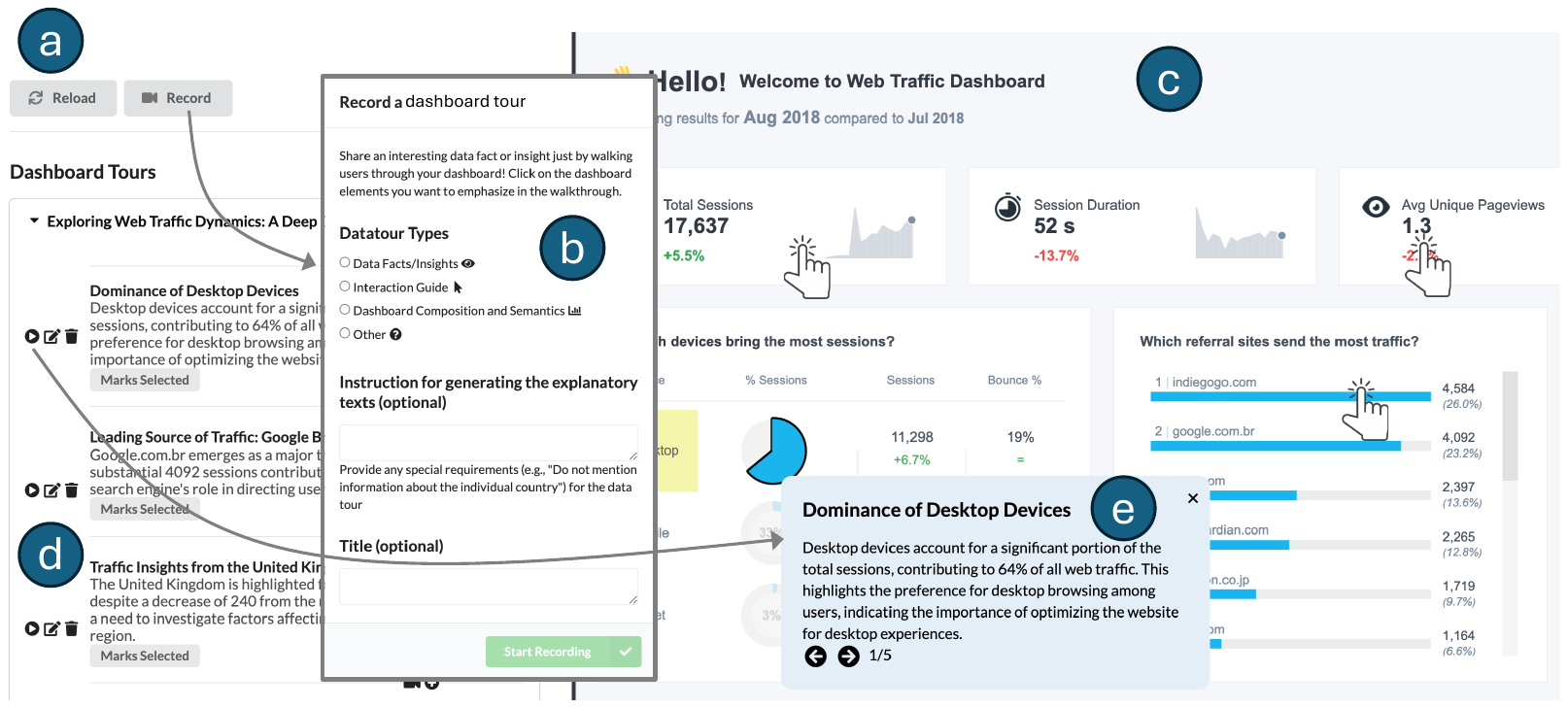}
    \caption{\textbf{Visual Interface of \toolname{}.} This example uses the same dashboard as the teaser image but the communication goal of the dashboard tour is \textbf{data facts} instead of \textbf{dashboard semantics and encoding}. a) Buttons for recording interactions and reloading a dashboard. b) On clicking the Record button (\faVideo), a modal opens where authors can select a communication goal, provide instructions to the LLM (optional), and a title (optional). c) After choosing the communication goal, authors can interact with the dashboard for creating the tour (shown with the hand cursors). d) Based on the interactions, a dashboard tour is created using an LLM. Each step in the tour contains three icon buttons for playing back the original interaction, editing the step, and deleting the step.  e) Playback of a step in the dashboard. The title and description appear as an overlay in the dashboard.}
    \label{fig:interface}
    \vspace{-1em}
\end{figure*}

\subsubsection{Guidance communication goals can vary}
\label{sec:formative-findings-intents}
\cre{Based on participant feedback, we identified four different kinds of communication goals for guidance focused on prescribing.} The first kind is \textit{dashboard semantics and encoding}, which explains the different elements of a dashboard, e.g. chart encodings (P2, P4, P6) or purpose of a dropdown menu (P2, P7). The second one is \textit{interaction guidance}, focusing on explaining possible ways to interact with charts and their impacts on other dashboard components (P1-9). The third one is \textit{data facts}, which shows insights and interesting data facts (P1-5, P8-9). Finally, \textit{data backgrounds and context} explains background information and context about the dashboard (P1-3, P8). 
\cre{These findings align with prior research, which identified similar guidance intents~\cite{DBLP:journals/tvcg/CenedaGMMSST17, DBLP:journals/tvcg/SultanumText2024}. For instance, \textit{dashboard semantics} and \textit{interaction guide} overlap with the VA methods and infrastructure categories proposed by Ceneda et al.~\cite{DBLP:journals/tvcg/CenedaGMMSST17}, both of which focus on explaining charts, widgets, and dashboard infrastructure. Similarly, \textit{data facts} overlap with three categories proposed by Ceneda et al.: \textit{data, task, and users}. }

\subsection{Design Goals}
The interviews suggest authoring effective guidance to prescribe to users is a time-consuming task. Participants were positive towards dashboard tours to prescribe guidance but had limited tools available to author them. We also identified four different communication goals for prescribing guidance. Based on the interviews, we decided to focus on mediating effectiveness and efficiency to facilitate the authoring process for dashboard tours. 
We coalesce the findings into a list of design goals (\textbf{DG}), as follows.

\textbf{DG1: Expedite authoring while supporting expressivity and ownership.}
Dashboard tours should be easy to author, while authors should still retain ownership of the content. Authors should be given quick starting points and reasonable defaults as well as the  ability to easily iterate over style and tone. \cre{D-Tour~\cite{dhanoad} is designed on a similar principle, but offers only template-based content generation. Our tool could use more advanced technology, such as LLMs, to expedite authoring and broaden expressivity in guidance. LEVA~\cite{leva} showed that LLMs could be useful in generating onboarding materials. However, it does not offer any features to integrate author oversight. It also does not focus on dashboard tours. Our tool should address this gap by using a human-in-the-loop approach to integrate LLMs for dashboard tours.}

\textbf{DG2: Support communication goals.}
\cre{Past work~\cite{DBLP:journals/tvcg/CenedaGMMSST17}} and formative findings point to a range of relatively well-defined communication goals \cre{(Sec.~\ref{sec:formative-findings-intents})}, which could be useful as guard rails for content generation while maximizing the utility and relevance of generated content. This range of goals should be acknowledged and supported, while still empowering authors to redirect goals as needed. \cre{While prior theoretical research has reported on the different goals~\cite{DBLP:journals/tvcg/CenedaGMMSST17, DBLP:journals/tvcg/SultanumText2024}, no prior tools explicitly support these communication goals~\cite{dhanoad, leva}.}


\cre{\textbf{DG3: Embed guidance into the dashboard.}} A critical decision for a dashboard tour is the format and content of an individual step. \cre{We found that dashboard tours should offer explanatory, data-relevant text in each step as an \textit{embedded} dashboard element (Sec.~\ref{sec:formative-authoring-guidance}).  Prior research also found embedded elements, often as overlays, to be effective at providing contextual guidance~\cite{dhanoad, DBLP:conf/vinci/StoiberWGSSA21, DBLP:conf/chi/NetworkNarratives}.} 

\begin{figure*}
    \centering
    \includegraphics[width=0.75\textwidth]{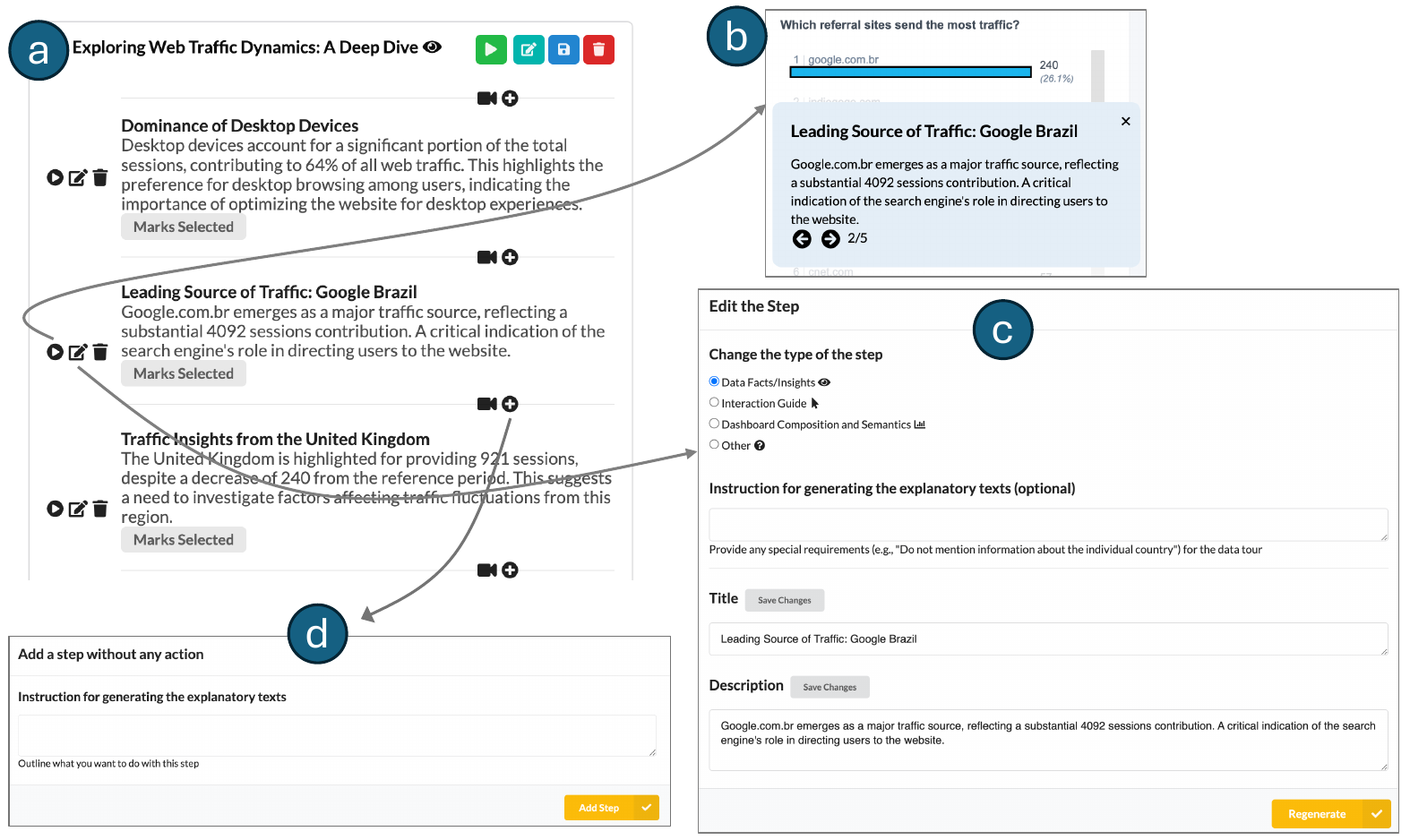}
    \caption{\textbf{Interfaces for editing a tour.} a) The top row in the interface shows the title of the tour as well as four editing features for the tour. The play button (\faPlay) hides all other components in the tool, except the dashboard, and then plays all steps one by one. The edit button (\faEdit) opens the setting modal from ~\autoref{fig:interface}b. The other two buttons allow saving (\faSave) and deleting (\faTrash) the tour.
    Each step of the tour contains three icon buttons. b) The first button (\faPlayCircle) plays back the interaction associated with the step. c) The second button (\faEdit[regular]) opens a modal for editing the setting for an individual step. This interface allows users to change the guidance intent for a step, provide further instructions for regenerating the step, and manually edit the title and description. d) Users can insert a new step without any interaction by clicking the plus icon (\faPlusCircle) and then outlining the instruction (e.g., \textit{``add a transition step here''}) for the step. Authors can also click on the record buttons (\faVideo) in between the steps to add new steps with interaction. }
    \label{fig:tour_interface}
    \vspace{-1em}
\end{figure*}
\section{DashGuide}
\toolname{} is a web-based and LLM-assisted proof-of-concept system to support authoring tours for \cre{Tableau dashboards} (\autoref{fig:interface}). We chose Tableau \cre{due to its prevalence for} dashboard authoring, the large community of publicly available dashboards in \cre{Tableau Public~\cite{tableau-public}, and its} API that allows for Tableau dashboards to be embedded into web-based applications~\cite{tableau-embedding}.

\subsection{Extracting Dashboard Metadata}
\label{sec:extract}
To expedite authoring, \toolname{} automatically extracts several metadata about a dashboard using the Tableau Public API (\textbf{DG1}). This process results in dashboard metadata about charts, images, UI widgets (e.g., dropdowns, filters, etc.), and text content. For instance, for a chart, we extract its position, encoding, and marks. For a text object, we extract the raw text data and its position in the dashboard. We store this metadata in a JSON object and feed it as context to the LLM. \cre{More details about the metadata and JSON object are available in supplemental materials.}

\subsection{Recording Dashboard Tour}
\label{sec:record_interactions}
To outline the steps for a dashboard tour, an author needs to follow two steps: 1) selecting tour metadata; and 2) interacting with dashboard components. This allows the author to craft the skeleton of the tour within minutes and with only a few user actions (\textbf{DG1}).

\subsubsection{Tour Metadata}
\label{sec:global_setting}
\autoref{fig:interface}b shows the interface for providing metadata about the tour. The primary purpose of this step is to collect the author's communication goal for the tour (\textbf{DG2}). The collected metadata are:

\begin{itemize}
    \item \textbf{Communication Goal.} \toolname{} offers authoring support for three different types of communication goals: \textit{dashboard semantics, interaction guide,} and \textit{data facts} (\textbf{DG2}). The goal drives the type of content for the data tour. For instance, for a tour with \textit{dashboard semantics} as the goal, we provide the following definition to the LLM: \textit{``Focus on explaining chart encoding and markers and the purpose of different filters, dropdowns, and other UI widgets in this dashboard tour.''}  For \textit{interaction guide} the prompt is: \textit{``Focus on explaining interaction and the effects of interactions on other coordinated views for this dashboard tour.''} 
    
    \item \textbf{Instruction (optional).} Authors can specify any custom instructions, requirements, or preamble for the tour, such as\textit{``Write the tour in third-person point of view.''}.
    
    \item \textbf{Title (optional).} Authors can provide a title for the tour. If provided, \toolname{} will use \cre{it} as a context for the tour. If not, the tool will prompt the LLM to generate a title for the tour.
\end{itemize}

We refined the prompts for the metadata iteratively, informing several design criteria. For example, we noticed GPT-4 was not producing correct descriptions for the \textit{data backgrounds and context}, one of the categories identified from the formative study (Sec.~\ref{sec:formative-findings-intents}). This was because a large portion of the information on data typically is hidden in a Tableau Public dashboard for security reasons. For the sake of tapping from the large collection of real-world dashboards Tableau Public offers, and to avoid the incorrect LLM generations, we decided to forgo this category in this proof-of-concept implementation of \toolname{}. We also experimented with mixing multiple communication goals for a tour. However, the generated content did not reflect the multiple goals correctly in most cases. Thus, we decided to constraint a tour to one goal although authors can change the goal of an individual step.

\subsubsection{Capturing Interaction}
After providing the metadata, an author can interact with the dashboard to create steps for the tour.
\toolname{} tracks \cre{common dashboard interactions~\cite{DBLP:journals/tvcg/YiKSJ07, DBLP:conf/vl/Shneiderman96}, including \textbf{data selection, brushing,} and \textbf{filters} applied to other views as a result of these interactions.} We do not track hovering actions: this is partly due to a limitation of the Tableau embeddings API, but also because
hovering interactions could create unnecessarily (and unintentionally) long interaction sequences with a large number of steps that authors would need to retroactively delete. We also track interactions with different UI widgets. These widgets include \textbf{buttons, dropdowns, checkboxes, and radio buttons}. We record changes to multiple coordinated views because of the interactions with the UI widgets. We store the interactions in a JSON object. We provide details about the JSON object in the supplemental materials.

\begin{figure*}
    \centering
    \includegraphics[alt={A figure with five sub-figures. Each sub-figure has a data visualization and a black rectangular box with some text, placed on top of the data visualization.}, width=0.95\linewidth]{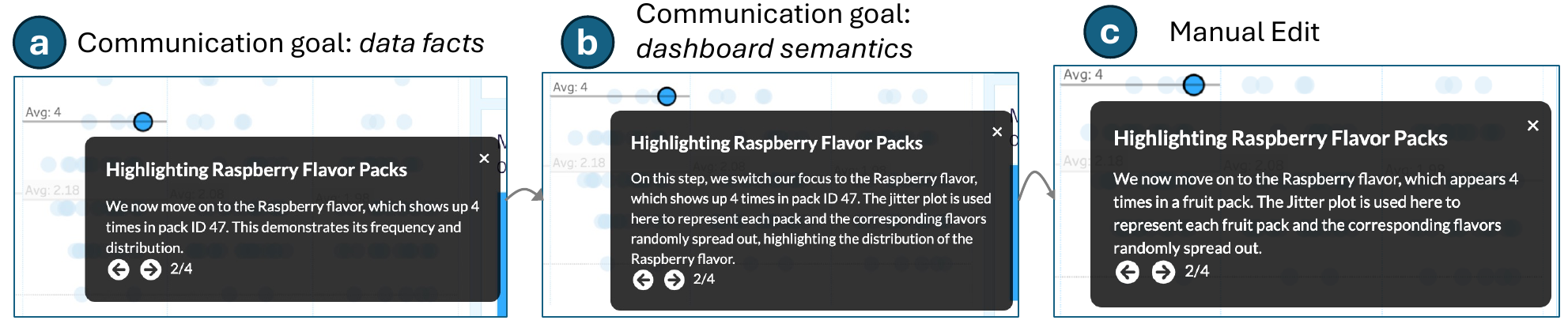}
    \caption{\textbf{Editing a Dashboard Tour in \toolname{}}. The dashboard in this example analyzes flavors \cre{(e.g., strawberry, peach, etc.)} in fruit snacks. Consider that the author \cre{created} a dashboard tour with the goal of communicating data facts. a) A step of the tour features a jitter plot. Thinking the users may not know about jitter plots, the author \cre{regenerates the step with another  communication goal, \emph{dashboard semantics}}. b) The step is updated with an explanation on jitter plots. c) The author manually \cre{tweaks} the description for a clearer explanation.} 
    \label{fig:example}
\vspace{-1em}
\end{figure*}

\subsection{Generating Tour}

\cree{As mentioned in Sec.~\ref{sec:extract}, we first package the metadata as a JSON and pass it to GPT-4 as context within a first prompt.}
We then prompt the LLM with the author-provided metadata and interactions as a second prompt to generate the initial dashboard tour. 
\cree{We iteratively refined the prompts by experimenting with various Tableau dashboards, all listed in the supplemental materials alongside our refinement strategy.}
\cree{}
The final prompt is as follows:

{\footnotesize \textit{``The following is a JSON object outlining the structure of a dashboard tour - }<a collection of interactions>\textit{. The key `steps' includes a list of author-performed interactions/events. The `coordinatedViewChange' key in each event represents changes to the coordinated views of the dashboard as a result of the event. The author also provided some metadata about the tour including a} <communication goal>\textit{.
Generate a title and short description for each event/step in natural language. Craft a story out of the steps and write the title and description accordingly. Return in the following JSON format \{'title':} <tour title>\textit{, steps: \{}<collection of event object with two keys, `title' and `description'>\textit{\}.''} }



\subsection{Editing Dashboard Tour}

Once the initial tour is created, it is crucial that authors are able to verify the content and revise it as they wish (\textbf{DG1}). To support that, \toolname{} offers several editing features. The features include:


\begin{itemize}
    \item \textbf{Editing Tour Metadata (\faEdit).}  Authors can use the interface discussed in Sec.~\ref{sec:global_setting} to revise the global settings of the tour.
    
        \subitem -- \textbf{Communication Goal.} Authors can change the communication goal (e.g., from \textit{data facts} to \textit{interaction guide}) to change the overall content of all steps within the tour. 
        \subitem -- \textbf{Instruction and Title (optional).} Authors can provide custom instructions (e.g., \textit{make the descriptions shorter}) to the LLM. Authors can also manually edit the title of the tour.
    
     \item \textbf{Save Tour (\faSave).} Authors can save the tour in our database.
    \item \textbf{Delete Tour (\faTrash).} Authors can permanently delete the tour.

    \item \textbf{Edit Content and Setting for the Steps.} Authors have control over individual steps.
    
        \subitem -- \textbf{Edit Title and Description.} Authors can use the interface in~\autoref{fig:tour_interface}c to manually edit the title and description for the step.
        \subitem -- \textbf{Change Communication Goal and Instruction.} Using the same interface, authors can change the communication goal or add new instruction for the LLM and regenerate the step.
        \subitem -- \textbf{Add New Steps (\faVideo).} In between the steps, there are recording buttons that the author can use to record new interactions and add them as new steps to the guide.
        \subitem -- \textbf{Add a Step Without \cre{an Associated} Interaction (\faPlusCircle).} \cre{This feature} helps authors insert \cre{standalone} steps that provide context, summary, or transitions to the narrative. 
    
\end{itemize}

The dashboard tour, including all the editing features, is encapsulated in a UI card (\autoref{fig:tour_interface}). We show an example scenario with different editing features in~\autoref{fig:example}.

\subsection{Delivery/Playback}
Following \textbf{DG3}, each step for a tour is delivered as an overlay text near the place of interaction (\autoref{fig:example}). Authors can also freely move the overlay to readjust its position. 

Authors can verify the delivery in two ways. First, they can use the play button with a step to play the particular step in the authoring view (\autoref{fig:tour_interface}b). Authors can use the forward and backward icons on the overlay to move back and forth in a tour. The other option is using the play button (\faPlay) at the top of the tour (\autoref{fig:tour_interface}a). This feature will hide all authoring controls on the interface and only show the dashboard and tour. This feature helps authors get a sense of how the tour will appear to an end user.


\section{Evaluation}
\label{sec:findings}

Following the design and development of \toolname{}, we conducted a user evaluation to validate its proposed approach for authoring dashboard tours. Participants used the tool to author tours for two dashboards (including one of their own, for those with Tableau Public profiles) and to share their impressions on the experience. We outline our study methodology and present our findings. 

\subsection{Recruitment and Methodology}

We recruited 12 dashboard creators with Tableau experience\cre{\,---\,P3 and P6 from the formative study (Sec.~\ref{sec:formative-participants}), plus ten new participants P10-P19\,---\,}via special interest groups. 
Participant backgrounds span data analysts and engineers (8), data consultants (2) and managers in relevant data/BI roles (2)\cre{, all with at least five years of professional dashboard authoring experience}. \cre{Additional participant details can be found in supplemental materials.}

Sessions were 1 hour long, conducted via video-conferencing with an experimenter and a note taker. Participants accessed \toolname{} on their own browsers while sharing their screen and thinking aloud. They performed two authoring tasks, each 8 minutes at most. Tasks were preceded by a demonstration of the tool, and were each followed by a brief discussion to reflect on participant's impressions and experiences.
For the 1st task, authors were given a dashboard out of a list of two dashboards (counterbalanced across participants) to create a dashboard tour focusing on a given guidance goal (between \textit{data facts}, \textit{interaction guide}, and \textit{dashboard semantics}, also counterbalanced across participants). This setup ensured a diversity of content generation exposure, but also a degree of consistency across experiences. After the 1st task, participants were asked for their impressions on the authoring experience, how they envisioned this tool fitting into their current practices, and alternative ways that guidance may be provided to their users.
For the 2nd task, 10 participants created a dashboard tour for a Tableau Public dashboard that they authored, and 2 used a Tableau Public dashboard they were familiar with. This helped provide more ecologically valid context for authoring. Follow-up questions focused on perceived time or quality gains with a tool like \toolname{} in their practice, and suggestions for improvement. \cre{More details on the study protocol can be found in supplemental materials.}

Sessions were screen-recorded, audio-recorded, and automatically transcribed. Extensive notes were also taken, and supported the initial stages of thematic analysis along with the transcripts. Usage logs were collected to inform basic usage patterns, including the first and last stages of their authored tours.  Qualitative findings drive most of the analysis, contextualized by illustrative quotes, relevant log stats, and frequency counts in the format (\_/12) denoting the number of participants who supported the theme. Frequency counts reflect \textit{spontaneous} comments from participants and \cree{are thus a \textit{lower bound}, and not a full signal,} of theme relevance.

\subsection{Findings}

Overall, participants reactions to \toolname{} were very positive. They appreciated the ease of use (5/12), and how quickly they can craft tours with it (5/12). When asked about foreseeable gains if using \toolname{} in their practice, most (11/12) concurred on at least one aspect of efficiency and quality, e.g., \qte{it would add value for the same time invested}{\cre{P15}}, and \qte{guidance is the last mile of dashboarding, the quicker I can make it the better}{\cre{P13}}. A majority of participants also commented on the higher-than-expected quality of content generation (7/12), e.g., \qte{I was pleasantly surprised when it did what I wanted it to do}{\cre{P14}}, and \qte{I like this much more than I thought I would}{\cre{P13}}.

Participants remarked on the importance of interactive and relevant hands-off guidance for dashboard users, as
\qte{there's that group of people for whom data is extremely overwhelming. And giving this to someone (..) this feeling of `someone has guided me around this insight', I feel like it would be awesome}{\cre{P11}}. Also, \qte{having this as a dashboard guide is super helpful, because it's a way to cut back on [users] needing me to go through things over and over again}{\cre{P6}}. Beyond being \qte{really helpful in training new users}{\cre{P16}}, one participant argued \toolname{}'s usefulness may extends to experienced users as well: \qte{maybe you haven't used the dashboard in a long time, you come back to it and you want to re-orient yourself to it}{\cre{P11}}.

Out of all participants, only one\,---\, who had significant experience in recording instructional videos for guidance\,---\, saw limited value in \toolname{}'s assistance: \qte{it has to be easier to use than any default video recording software (..) I was expecting the tool to do something similar [to a video recording], but it's more of a screenshotting tool.} {\cre{P17}}. 
On that, even other participants who were enthusiastic about \toolname{} also noted several points of friction, including usability and workflow issues. More notably, they identified contextual challenges beyond the current scope of \toolname{} that point to compelling research opportunities for future work, which we cover later in Sec.~\ref{sec:findings-oppties}.

Here, we discuss strengths and opportunities for improvement for \toolname{}. We organize our findings under four themes: observations on authoring workflows (Sec. \ref{sec:findings-workflow}), impressions on AI-generated content (Sec. \ref{sec:findings-generation}), reactions to \toolname{}'s approach to communication goal (Sec. \ref{sec:findings-intent}),  and comments on formatting and delivery of guidance to end users (Sec. \ref{sec:findings-guidance}). 


\subsubsection{On Authoring Workflows and Interaction Capture}
\label{sec:findings-workflow}


Findings suggest that even as a proof-of-concept, \toolname{}'s interaction-centered workflow not only worked well as a tour authoring approach but also supported a diversity of authoring workflows.
Within the allotted 8 minutes for each task, participants were able to quickly internalize and appropriate the authoring features. 
After the initial recording, \cre{on average, participants appended 1.50 (CI: 0.00, 3.09) new interactive steps and 2.20 (CI: 0.58, 3.82)  new non-interactive steps (see supplemental materials). On average, participants re-generated 2.00 (CI: 0.52, 3.48) steps with the help of LLM and manually edited 3.33 (CI: -6.71, 13.37) steps. Finally, participants regenerated the full tour 2.43 (CI: 0.30, 4.55) times on average. A chart is available in supplemental materials.} 


Some participants found \toolname{}'s interaction-centered approach to be intuitive, both from a guidance perspective, e.g., \qte{I want to be able to show them exactly where to go}{\cre{P6}}; and from a usage perspective, e.g., \qte{it makes sense for me to do the action that I think the user will take}{\cre{P10}}.
Many also stated that authoring flexibility is important to them 
and praised features of the tool that support this freedom, e.g., inserting steps in the middle (\cre{P6}), re-prompting (\cre{P11}, \cre{P14}), and direct editing (\cre{P13}). This diversity underscores the \cre{wide} range of workflows \toolname{} supports. 


Participants also highlighted limitations with \toolname{}'s proposed workflow. First, it requires an initial sequence of interactive actions before content is generated, which imposes a higher degree of deliberation and planning from the author and is thus less amenable to narrative and flow exploration. 
\cre{P14} also anticipated challenges arising with more spontaneous interaction workflows: \qte{I would find that having to delete stray marks, because I click around too much, would annoy me after a while. Especially if I was going through multiple multi-step tours.}{\cre{P14}}. 
Suggested improvements include pre-filtering interactions of interest, means for easy editing of captured step sequences before generation, and the ability to group steps into sections or segments.

\subsubsection{On Content Generation}
\label{sec:findings-generation}


Impressions on content generation were generally positive, with 7/12 spontaneously stating they appreciated the auto-generated assistance, including self proclaimed ``skeptics of AI'': \qte{I'm typically quite skeptical of everything that has AI baked into it, but this works}{\cre{P15}}. One benefit they found was expedited authoring: \qte{`Data facts' and `dashboard semantics' are the most labor intensive. I could see that saving a lot of time, a lot of headache, a lot of maintenance work, and a lot of repeated education}{\cre{P18}}.
Another benefit was helping overcome the \qte{blank screen}{\cre{P14}} paralysis: \qte{These are things that I would probably spend a lot of time writing up and thinking about and I don't feel like I'd have to make a lot of changes.}{\cre{P16}}. 
These findings are echoed by think aloud utterances, where 10/12 participants shared positive reactions toward content generation at least once across the two tasks. To many, it largely fulfilled expectations from the get go, as about 72\% of all authored steps were kept as is: e.g., \qte{the insights that came out, right out the gate, I either didn't have any edits, or the edits that I did have were very easy}{\cre{P18}}. This suggests our proposed AI-assisted workflow is helpful and that content generation does bring some efficiency. 
Some also liked the responsiveness to custom instructions, e.g., \qte{it was about 50\% but changing that prompt got it to about 90-95\% of where I'd want it}{\cre{P6}}.
Some also appreciated the external knowledge brought in, e.g., the \qte{extra context for the analysis that I do not provide on this dashboard}{\cre{P12}}. 


Regarding opportunities for improvement, suggestions called for more implicit and explicit context management over generative AI. The most common ask was having more control over context to be passed to the large language model (5/12). Participants wanted specific information to be taken into account, e.g., recent or specialized knowledge, which could be potentially mediated by data dictionaries and shared semantic layers. But when internal labels used in the Tableau workbooks appeared in the content, some participants wanted them \textit{excluded} from the generation context as they were not always user friendly: \qte{what we name our worksheets probably isn't relevant to users}{\cre{P16}}. One participant (\cre{P3}) suggested interpreting other visual dashboard elements, e.g., a magnifying glass for a search icon, and factor their meaning into the generation. 
Other ideas included more proactive interventions over content (3/12), e.g., suggesting initial steps for an \textit{interaction guide} tour based on interactive widgets present on the dashboard (\cre{P3}), and recommending follow-up steps to append to existing tours~(\cre{P11}).

AI skepticism aside, several participants (4/12) working in  data-sensitive domains (e.g. governmental) also shared practical obstaclesto the use of large language models in their dashboard authoring practices, from security restrictions to  high cost and sustainability concerns. As such, they hoped for an option to \qte{skip the AI}{\cre{P13}}, or potentially have it replaced by classic template-based approaches (\cre{P16}), arguing \toolname{} provides value even without the generative piece: 
\qte{If you were to launch this annotation experience without ChatGPT integration you would still have huge gains, because you can enter your own insights. The ability to create an annotated walkthrough in the manner you created is a huge win}{\cre{P11}}.


\subsubsection{On Communication Goal}
\label{sec:findings-intent}


Participants chose a variety of communication goals on their free form task (2nd task), including 6 \textit{data facts}, 4 \textit{interaction guides}, and 2 \textit{dashboard semantics}. Feedback points to these types being largely aligned with communication needs of dashboard authors: \qte{I really think the 3 categories are 90\% of what an annotation guidance would even need}{\cre{P18}}. 
Several participants (4/12) also appreciated \toolname{}'s potential for storytelling beyond guidance, such as presentations (\cre{P14}). Overall, we found the \textit{communication goal} approach to be an effective and versatile authoring scaffold, giving authors a range of guidance options while also equipping content generation with suitable guard rails.


As next steps, several participants (5/12) suggested expanding the list of communication goals to support \textit{data backgrounds and context}, the specific communication goal we chose not support in \toolname{}'s current implementation (as covered in Sec.~\ref{sec:global_setting}) but is still feasible within our framework: \qte{In my day to day,  it's something that we're always asking: `what's the numerator', `the denominator',  `what's feeding into these 1\%', etc.}{\cre{P18}}. One participant also suggested experimenting with multiple active communication goals per tour: \qte{I imagine there being times I'd want to combine them and, instead of one or the other, create a guide that does all of them together}{\cre{P16}}.
Others also suggested having more control over global tone and wording choices (3/12), e.g., a slider for positive or negative tone (\cre{P10}). 

\subsubsection{On Guidance Delivery}
\label{sec:findings-guidance}


\toolname{}'s embedded step-by-step format was informed from our formative investigation presented in Sec. \ref{sec:formative-authoring-guidance}. In this evaluation, participants echoed similar impressions; they 
praised \toolname{}'s co-located delivery (3/12), how actions are replayed while the dashboard is still interactable (3/12), and the styling of content pop-ups (4/12). The ability to create multiple data tours for a dashboard was also acknowledged (3/12) and considered useful.


Other priorities participants had for guidance delivery include customizing text and pop-up styling (6/12), but also considering styling in the context of accessibility, for things like font size and contrast (2/12). On that, a text-based delivery may be considered a potential boon for accessibility provided that \qte{screen readers can tap into those blurbs}{\cre{P6}}. Other opportunities included consistent entry points in dashboards for triggering tours (\cre{P6}); flexible user controls to choose between multiple tours or from which step to start the tour (\cre{P10}); multi-modal presentation, e.g. with video excerpts and animated \textit{gifs} (\cre{P6}, \cre{P19}); and a user-accessible table of contents (\cre{P10}). Finally, some participants underscored the value of \toolname{} as a more general authoring tool and wished to see it applied to formats such as reports (\cre{P12}) and presentations (\cre{P14}).

\section{Discussion}
\label{sec:discussion}

Study findings suggest \toolname{} is both an effective and an efficient alternative for authoring dashboard tours. Participants were generally enthusiastic about the prospects of assisted authoring, and were often positively surprised by the generated content. This suggests that the guard rails we designed, which in turn were informed by feedback from practitioners and prior work, are well aligned with dashboard authors' communication needs. 
In the following segments, we take a step back and discuss opportunities for further investigation that emerged from our evaluation. 
We also reflect on the impact of this work in the space of dashboard guidance and LLM-assisted authoring, and discuss limitations of our work.


\subsection{Open Ended Challenges and Opportunities}
\label{sec:findings-oppties}

Study point to \toolname{} addressing a relevant need, but also informed areas for improvement. Some of these areas touched on open research questions, which could in turn inform compelling opportunities for future work. We discuss them as follows.

\subsubsection{Handling data updates}
\label{sec:findings-oppties-data-updates}
In our formative study, dashboard updates were considered a significant pain point for guidance maintenance. In line with this, one of the most frequent questions participants asked this time around was whether \toolname{} handled data updates (6/12), which naturally leads to the question of how generative AI may reconcile those changes with the guidance content. While generative AI could potentially update this guidance automatically, it should not do so without oversight, and the ideal extent and format for such oversight are not immediately clear.
Simple solutions exist, such as automatic alerts on data changes (\cre{P13}), 
but additional guard rails to improve quality of generated content would be worth investigating. 
Another solution is to combine generation with template-based approaches and text-data links, such as the dynamic text placeholders often used in dashboard tooltips and data labels~\cite{DBLP:journals/tvcg/SultanumText2024}: \qte{I would expect to be able to use field references in the descriptions.}{\cre{P14}}





\subsubsection{Reusing guidance content}
\label{sec:eval-findings-reuse}

One participant remarked on the opportunity to reuse previous tours: \qte{If you've got dashboards that are very similar, it would be really nice to reuse a tour.}{\cre{P10}}. Following up on Sec.~\ref{sec:findings-oppties-data-updates}, previous content could serve as guard rails to generate new guidance content, constituting a form of reuse. 
Other reuse scenarios could include repurposing tours for different target audiences (e.g., sales and marketing"), by following the same interaction route but adapting the language and takeaways; 
collectively repurposing tours from the same source (e.g., an author or team) for structure and tone, to create new content that is aligned and consistent with past content; 
and having end-users reapplying style and interaction flow from past guidance they liked onto a new dashboard.


\subsubsection{Adapting to users}

Supporting audiences with different backgrounds, goals, and skills is one of the many challenges guidance authors face. \toolname{} enables that via rapid authoring of multiple tours for a dashboard, which is often a desirable setup,  but still requires authors to proactively create tours ahead of time and know their users well: \qte{I don't want business users from this department looking at [a dashboard] the same way as the accountants, especially if I am trying to use a single dashboard to serve multiple purposes}{\cre{P14}}. Beyond further automation via content reuse (Sec.~\ref{sec:eval-findings-reuse}), there is also an opportunity to tweak guidance at delivery time. One idea is to pick up contextual signals from the user (such as a trace of dashboard interaction) to identify relevant guidance steps from existing tours and selectively present them. Another idea is to use generative AI to create transitions between steps, adding cohesive flow while still largely respecting pre-vetted guidance content. Finally, one participant argued collective interaction traces from many users could also provide useful signals to guide dashboard tour authoring:\qte{I'd like to see where all those clicks are distributed}{\cre{P12}}. 


\subsubsection{Building on the exploration-explanation continuum}

In view of \toolname{}'s applicability to storytelling-related guidance, we argue it could also be used to chronicle an analytical exploration workflow performed on the dashboard. This idea is inspired by prior research in provenance of user interactions for sensemaking~\cite{DBLP:journals/cgf/XuOWSCW20} and mediating sensemaking and communication~\cite{DBLP:journals/tvcg/ChenLAAWNT20}, as well as a participant's comment on early stages of authoring: \qte{when you're in an authoring flow, when you are clicking somewhere, you don't know whether the insight you discovered is gonna be worth sharing. You are not ready to add it to your tour yet, but you think it might be. (..) You want to add bookmarks in the flow of analysis of something that you might want to refer to later}{\cre{P11}}. On this, \toolname{} could be extended with means to retrieve, organize, regroup and contextualize interaction traces that can be flexibly mapped dashboard tour steps later.



\subsection{Reflections on the Authoring Workflow in \toolname{}}
Despite the populatity of prompt-based interaction, they can also be misaligned to how humans perceive tasks and goals~\cite{DBLP:conf/iui/KimL0PK24}. Subramonyam et al.~\cite{DBLP:conf/chi/SubramonyamPPAS24} discussed the case of ``gulf of envisioning'', or the challenges of tweaking a prompt to improve output quality. In \toolname{}, authors prompt the LLM through a mix of UI widgets, dashboard interactions, and direct instructions, thus deconstructing the prompting task into smaller visual tasks. We believe this approach helped reduce the ``gulf of envisioning'' for authors. We hope our work, along with other recent works in HCI~\cite{DBLP:conf/chi/ArawjoSVWG24, DBLP:conf/uist/0002LCK23}, helps motivate future research to consider visually constructing prompts with smaller tasks.

\subsection{Limitations}
\toolname{} has several limitations. The tool is built upon Tableau Embedding API, which allowed us to concretely evaluate the tool on a very diverse ecosystem of real-world dashboards, but with limited support for capturing interactions (e.g., hovering actions). \cre{Nevertheless, the \textit{interaction-first} approach proposed in this work is platform-independent and could be implemented for other platforms by following the structure of the JSON files and LLM-prompts. Detailed instruction for adapting \toolname{} to other platforms can be found in the supplemental materials.}

In its current form, \toolname{} lacks a few post-capture editing features, such as merging steps.
\cre{We also acknowledge that mapping one step per one interaction may not capture complex insights, generated by a sequence of interactions. }Similar to other LLM-based applications, there is a chance of the LLM generating incorrect or non-existent information (i.e., hallucinations)~\cite{DBLP:journals/corr/abs-2108-07258}. Finally, while our results show promises about how end-users might benefit from the tours, we have not formally evaluated \toolname{} with \cree{them, something which would help characterize the extent of \toolname{}'s benefits and better assess its delivery format.}





\section{Conclusion}
We presented \toolname{}, a framework and system that can support dashboard creators in authoring interactive dashboard tours for guiding end users. 
This paper advances the state-of-the-art for dashboard tours by demonstrating how we can utilize \cre{LLMs to flexibly author the content of the tours}. We conducted formative interviews to identify design requirements for LLM-assisted authoring of dashboard tours. 
Based on author feedback, we designed different aspects of the tool, from communication goals to guidance delivery and authoring control over the content of the tours. We conducted a summative user evaluation with 12 dashboard creators, where they authored guided tours for existing dashboards using our tool. \toolname{}'s interaction-first authoring framework was found to be efficient and intuitive, which we argue may lower barriers for \cre{authoring} dashboard guidance. We also identified challenges and opportunities for future investigations to guide further research on the topic. 
As a community, we are still in the early stages of learning how to utilize this technology in ways that uphold human agency, awareness, creativity, and trust. We hope our work can help inspire other LLM-assisted work striving for the same goals.



\bibliographystyle{eg-alpha-doi}  
\bibliography{references}        

\end{document}